\documentstyle[12pt]{article}
\begin{document}
\newcommand{\be}{\begin{equation}}
\newcommand{\ee}{\end{equation}}
\begin{center}

{\bf Charm in cosmic rays}\\
(The long-flying component of EAS cores)

\vspace{2mm}

I.M. Dremin\footnote{e-mail: dremin@lpi.ru}, 
V.I. Yakovlev\footnote{e-mail: yakovlev@sci.lebedev.ru}\\

{\it Lebedev Physical Institute RAS, 119991 Moscow, Russia}

\end{center}

\vspace{1mm}

\begin{abstract}
Experimental data on cosmic ray cascades with enlarged attenuation lengths
(Tien-Shan effect) are presented and analyzed in terms of charm 
hadroproduction. The very first estimates of charm hadroproduction cross 
sections from experimental data at high energies are confirmed and compared 
with recent accelerator results.
\end{abstract}

\section{Introduction and brief overview}

Charm was found in cosmic rays in 1971 by Niu et al \cite{niu} (no such name 
was ascribed to it at that time). The creation of charm particles provoked 
in 1975 to relate them \cite{niko} to the long-flying component of the cores of 
extensive air showers (EAS) observed a couple of years before \cite{ase, agnp} 
and named as the Tien-Shan effect. This idea was however abandoned for several 
years because no precise data on properties of charm hadrons existed. Let us 
note that the similar elongation of the cascades observed at Aragatz 
installation in Caucauses with rather low statistics was analyzed in late 1970s 
in \cite{dmsr} but the resulting estimates of particle parameters (mass about 
10 GeV and lifetime 10$^{-10}$ s) were misleading.

Open charm measurements in accelerator experiments date back to late 
1970s when $D$ and $\bar D$ mesons were first detected. In earlier 
1980s, the leading effect in $\Lambda _c$ production was declared by
experimentalists \cite{basi, bode} and supported by theorists \cite{khod}.
The small inelasticity coefficient for $\Lambda _c$ was also advocated
\cite{khod, mcl}. The spectra of $D$-mesons were considered to be much softer 
than those of $\Lambda _c$ \cite{bode, hwa}. The charm hadroproduction cross 
sections 
measured at energies $\sqrt s < 20$ GeV were quite small (less or about 10 
$\mu $b). It looked improbable that they increase fast with energy even though
first calculations in the quark-gluon strings model showed \cite{bkal} that they
can become as large as 0.1 -  1 mb at energies exceeding $\sqrt s =100$ GeV.
Smaller values were however obtained in \cite{ger}.

Meantime, masses and lifetimes of charm particles were measured more and more
accurately. In parallel, the more precise data about the long-flying component
of the EAS cores were obtained \cite{npes}. The specific structure in the energy 
dependence of the attenuation length of cascades observed in the hadron 
calorimeter \cite{yako, yak} revived the idea about charm production 
responsible for these peculiar features \cite{dyak}. The starting impact was
related with long lifetimes of charm particles. Both analytical and computer
calculations with kinetic equations \cite{dmsa} and Monte-Carlo simulations
of cascades in the calorimeter \cite{drya, madi, dima} were attempted. They 
lead to the conclusion \cite{dyak, drya, viy} that the charm production cross 
section can be as large as 1.4 - 2.8 mb at the laboratory energy $E_L= 10 - 20$ 
TeV ($\sqrt s = 140 - 200$ GeV) and about 3 - 5 mb at $E_L\approx 100$ TeV 
($\sqrt s \approx 450$ GeV). These estimates were the earliest values for the
charm hadroproduction cross sections obtained from experimental results at very 
high energies. The cosmic ray data obtained with lead and X-ray emulsion 
detectors also showed the existence of the long-flying cascades \cite{pami}. 
Later it was concluded \cite{ivan, rako} that they support a large charm cross 
section at $\sqrt s = 300$  GeV. These values of the charm hadroproduction cross 
section were considered as being extremely large until recent experiments at 
RHIC energy $\sqrt s = 200$ GeV claimed (see, e.g., \cite{dong}) that this 
cross section per nucleon in $pp$ and d-Au collisions is $1.4\pm 0.2\pm 0.4$ mb 
according to 
STAR collaboration \cite{star, sta2} and $0.92\pm 0.15\pm 0.54$ mb according 
to PHENIX collaboration \cite{phen}. The same collaborations extracted the same
cross sections from Au-Au collisions at $\sqrt s = 200$ GeV and give the 
following numbers: $1.11\pm 0.08\pm 0.42$ mb for STAR \cite{zhan} and 
$0.622\pm 0.057\pm 0.160$ mb for PHENIX. The compilation of both accelerator 
and cosmic ray data on charm hadroproduction cross section at various 
energies is demonstrated in Fig. 1.

The difference by a factor about 1.5 - 2 between the two collaborations at 
RHIC is related to problems in extracting the cross section values. They 
are obtained from a finite number of measured $D$ mesons in 
a particular decay channels by using many correction factors. Especially 
important are the extrapolations to the full phase space because of undetected
forward region and the role of other numerous unmeasured charm hadrons. Let us 
stress here that namely forward region is crucial in cosmic ray experiments.

The situation with QCD calculations is also not clear yet even though there 
has been a great deal of improvement over last 10-15 years. In 1990s they
predicted rather low values of these cross sections with a very mild increase
with energy. Recent results taking into account higher order (NLO) perturbative
corrections \cite{cnv, vogt} give larger values and improve the situation. They are
however still lower by a factor 5-6 than STAR data. The quark-gluon strings
model is in a better position (see, e.g., \cite{pisk, pnik}) predicting
larger cross sections. The heavy quark production was also considered in the
semihard QCD approach \cite{bzli}.

In view of this intriguing and rapidly evolving situation we decided to
reanalyze previous cosmic ray data and compare with results obtained during
last years.

Let us mention that the charm particle production is also important for  
muon studies, both underground and in gamma-astronomy \cite{dmv, zats}.

\section{Qualitative expectations}

Before discussing the experimental installation in detail, we would like to  
explain the physics of the phenomenon and present the qualitative expectations 
which gave rise to the idea about the role of charm particles. EAS cores
consist of beams of high energy hadrons. These hadrons interact actively when 
they pass through the dense matter of the calorimeter. Most interactions give 
rise to abundant pion production. These pions create new pions in inelastic 
collisions. The hadronic shower is developed with the typical attenuation length 
in lead about 600 - 700 g/cm$^2$ . In some events the charm particles are 
however produced. Their 
decay lengths are of the order of tens or hundreds $\mu $m at comparatively 
low energies. Thus the low energy charm particles decay within the main shower 
and nothing special happens. The decay length is proportional to the 
$\gamma $-factor. Therefore, high energy charm particles are able to penetrate
to larger depths in the calorimeter. If they carry large portion of initial 
energy, then the shower elongates and the attenuation length increases.
However, at a somewhat higher energy it can happen that a produced charm
particle passes through the whole calorimeter without decay. Its energy is no 
more detected in the calorimeter. The attenuation length should come back to 
its standard values. Thus one would expect to observe a maximum in the energy
dependence of the attenuation length.

Let us advertize in Fig. 2 the final result of the Tien-Shan experiment where
the observed energy dependence of the attenuation length of hadronic showers
in EAS cores is plotted. This anticipates its detailed discussion below. 
The upper and lower experimental points correspond to two classes of cascades
separated according to special methods among the available amount of data.

We will interpret the upper points as typical for long-flying cascades with
decaying high energy particles because they are obtained from samples enriched
by such particles as advocated below. With three peaks in it, one is
tempted to ascribe this plot to $D^{\pm }, D^0$ and $\Lambda _c$. The decay
length $l_i$ for any species $i$ is given by
\be
l_i=c\tau _i\gamma _i=c\tau _i\frac {E_i}{m_i},   \label{li}
\ee
where $\tau _i$ is its lifetime, $\gamma _i$ is the $\gamma $-factor,
$E_i, m_i$ are its energy and mass. If the interaction with production of
a charm particle took place in the upper part of the calorimeter, then the
attenuation length would return to its standard value at $l_i=z_c/\rho $ 
where $z_c$ is the calorimeter depth in g/cm$^2$ and $\rho $ its average 
density in g/cm$^3$. Then all particles with energies exceeding
\be
E_i=\frac{m_iz_c}{c\tau _i\rho}  \label{ei}
\ee 
slip down the calorimeter and decay outside it. 

Thus the peak can be ascribed only to those particles whose energies are 
high enough to decay at lengths larger than 0.3 - 0.5 m and lower than 
$E_i$ (\ref{ei}). Its shape and height are 
determined by the energy behaviour of charm hadroproduction cross section which
favours higher energies and drives the peak to (\ref{ei}). For the parameters of
Tien-Shan calorimeter $z_c=850$ g/cm$^2$, $\rho =3.54$ g/cm$^3$, i.e.,
$z_c/\rho =240$ cm, and particle masses and decay lengths 
$m_{D_{\pm }}= 1869$ MeV, $m_{D_0}=1864.6$ MeV, $m_{\Lambda _c}=2284.9$ MeV,
$c\tau _{D_{\pm }}=311.8$  $\mu$m, $c\tau _{D_{0 }}=123$ $ \mu$m, 
$c\tau _{\Lambda _c}=59.9$  $\mu$m one gets $E_{D_{\pm }}=14.4$ TeV,
$E_{D_{0 }}= 36.4$ TeV, $E_{\Lambda _{c}}= 91$ TeV. It follows that the initial 
energies should be high enough for such particles to be created \footnote{
The $\Lambda _c$ lifetime estimated in \cite{niu} was about 18 times less.
Therefore it required in initial estimates \cite{niko} too high energies 
that prevented further work on this hypothesis.}. They can be 
estimated by dividing the above numbers by the Feynman ratios $x_i=E_i/E_0$.
Qualitatively we can say that the values of 
$\langle x_i\rangle \approx 0.2 - 0.3$ would correspond to peaks positions
in Fig. 2. The ratios of energies at which peaks are positioned in Fig. 2
correspond quite well to the ratios of energies of various charm particles.
Thus these findings favour the hypothesis about charm particles initiating
the effect of elongated cascades. 

This is however an oversimplified estimate. Beside knowing the spectra of charm 
particles, i.e., $x_i$-distributions, one should use the energy and atomic 
number dependence of the charm production cross section on nuclei, the cross 
section of the interaction of a particular charm particle in a medium and its 
inelasticity coefficient, i.e., the share of energy spent by it in interactions 
with the calorimeter matter. The primary spectrum and composition of cosmic 
rays as well as the
secondary interactions in the cascade must be taken into account as well. Also, 
one should know at which depth in the calorimeter the charm particle was 
produced because in the proper estimate for $E_i$ the value $z_c$ should be 
replaced by 
$z_c - z_1$ where $z_1$ is just this depth. After all these factors are
accounted in Monte-Carlo simulations, the results and their sensitivity to
various parameters can be quantitatively studied. Nevertheless, one can start 
the next stage of qualitative approximations with analytical approach using 
the simplified kinetic equations. Before doing this we describe in more details 
the experimental installation and methods of analyzing experimental data.

\section{Experimental installation}

Extensive air showers (EAS) were registered in the ionization calorimeter 
placed in the Tien-Shan mountains at the altitude 3330 m. It consists of the 
lead absorber and ionization chambers of the total area of 36 m$^2$ and the 
height 2.4 m. The total thickness of the lead absorber (including some other 
elements recalculated to lead) is 850 g/cm$^2$. There are 17 rows of 48 copper 
ionization chambers with size $5.5\cdot 24\cdot 300$ cm$^3$. Each chamber has 
its own analogue digital convertor (ADC) with the dynamical range 
2$\cdot$10$^4$. The accuracy of signal measuring is 
better than 10\% in the whole range. Signals from the chambers are stored 
in the diode-capacitor cells. Then the series of pulses is sent to all ADC 
inputs. Each subsequent pulse is 10\% higher than the previous one. If the 
height of the pulse exceeds those stored by 10\%, its number $n$ is fixed. Thus 
all ADC are calibrated after each trigger. The information is stored in 
magnetic tapes.

The operation control of the calorimeter includes the everyday statistical 
analysis of each channel. The statistics in individual ionization chamber is
compared with the average statistics in each particular row.

\section{Methods of analysis of long-flying cascades}

The shower array selects EAS with the number of particles exceeding 
1.3$\cdot $10$^5$. Then the EAS cores which do not cross the calorimeter 
sides were selected for analysis.

In Fig. 3 we demonstrate an averaged cascade at the energy 37.6 TeV. The
averaging has been done over 765 events. It is clearly seen that the
electron-photon component of EAS dominates at the depths less than 133 g/cm$^2$.
Therefore, the energy of the hadronic component has been estimated by the energy
released in the calorimeter at the depths from 133 to 850 g/cm$^2$. The events
are selected according to this value.

The ionization curves of the averaged cascades are approximated by the 
exponential function $\exp (-z/L)$ in the depths interval 344 - 850 g/cm$^2$ 
(see Fig. 3) and the attenuation length $L$ is calculated. Its mean value is  
estimated both by simple averaging and from the distribution of inverse values
$1/L$ in individual cascades. Even though the difference between these estimates 
is small, their average value is chosen for further analysis. Its energy 
dependence has been found in two runs of measurements (with statistics 6976 and 
3617 events at energies higher than 3 TeV) done in the 6 years interval. The 
results fully coincide that proves the stability of the calorimeter operation.
Also, the position of the absolute cascade maximum is determined. It has been 
shown that it is directly correlated with the energy dependence of the 
attenuation length.

Moreover, it was attempted to find how the energy is distributed around the
cascade axis at different energies. For this purpose, the ratio $C(z)$ of the 
ionization energy released within the circle with the radius 36 cm around
the axis in the row of chambers at some depth $z$ to the total energy in all 
chambers of the same row has been measured. This method of analysis is crucial
for separation of cascades with charm particles from the standard cascades. 

\section{Results}

The most spectacular feature of the attenuation lengths energy dependence 
is their increase from values about 900 g/cm$^2$ at 20 TeV to about 
1500 g/cm$^2$ at energies close to 100 TeV and 1900 g/cm$^2$ at 300 TeV
with their decrease at even higher energies as
seen in Fig. 2 \footnote{The fine structure of this dependence is discussed 
below}. This increase is shown to be related to the shift of the 
absolute cascade maximum to larger depths with increase of the attenuation 
length (or energy) as demonstrated in Table 1.
\vspace{0.5cm}

{\bf Table 1}\\
The position of the absolute cascade maximum $z_m$ at different attenuation 
lengths $L$.\\
\vspace{0.5cm}

\begin{tabular}{cc}
\hline
 $z_m$, g/cm$^2$ & $L$, g/cm$^2$\\
\hline
 $374 $& $667\pm 15$ \\
\hline
 $600 $& $847\pm 60$ \\
\hline
 $>600$  & $2196\pm 260$\\
\hline
\end{tabular}

\vspace{0.5cm}
If this maximum is attributed to the decay products of a high energy particle,
it shows that this particle penetrates to larger depths in the calorimeter at
higher energies as one would expect for a particle with a fixed and rather 
large lifetime.

Further insight in the problem which helped reveal the fine structure of 
this increase seen in Fig. 2 was obtained from studies of the concentration 
behaviour. Its values have been measured for cascades with different 
attenuation lengths. The dependence of the concentration 
$C(z)$ on the depth $z$ was approximated by the simple linear line $C(z)=a+bz$. 
Actually, it has been observed that $b$ depends on the attenuation
length as shown in Fig. 4. Its average value at low attenuation lengths
$L<800$ g/cm$^2$ is negative $b_{low}=-1.15\cdot 10^{-4}$ cm$^2$/g but it is 
positive at large attenuation lengths $L>800$ g/cm$^2$
($b_{large}=1.4\cdot 10^{-4}$ cm$^2$/g). This demonstrates that at low energies 
the cascade energy spreads from the center while at large energies it tends to 
be more concentrated near it. The strongly attenuated cascades have a relatively
large transfer into their peripheral region, whereas the long-flying cascades
show increasingly more ionization concentation at larger depth values in their
core region. This could be understood by the conjecture that energetic 
long-lived unstable particles are produced in the long-range cascades.
The same feature is confirmed by the difference in 
the dependence of the positions of the absolute cascade maximum (Fig. 5) on 
the absorber depth. For $L<$800 g/cm$^2$ they decrease faster than for
$L>$800 g/cm$^2$: the exponents differ by the factor about 2.5.

These findings allowed to plot separately the energy dependences of $L(E)$ for 
cascades with $b>0$ and $b<0$.  The most remarkable difference between the two
new cascade subsamples appears in the energy dependence of the attenuation 
length (Fig. 2). For the cascades with the negative sign of $b$ the data are
compatible with the usual rather mild dependence shown by the lower dots in 
Fig. 2 typical
of single nucleon cascades containing only pions. This gives a confidence that
the proposed selection criterion according to the sign of $b$ properly separates
the standard cascades. As to the subsample of cascades with the positive values
of $b$, which is expected to be enriched with cascades containing the long-lived
unstable particles, the irregular behaviour with pronounced maxima is observed
(the upper dots in Fig. 2). All this favours the hypothesis that the decaying 
particle is the leading one in the production process. It plays more important 
role at high energies and releases its energy close to the shower axis.

The sign of $b$ has been used for estimates of the total cross section of charm 
hadroproduction. The individual cascades with energies higher than 10 TeV were 
separated according to the sign of $b$. It happened that they were split into 
two equal groups. The distributions of their attenuation lengths were obtained.
The cascades with positive $b$ are shifted to larger depths than those with 
negative $b$ (see Fig. 6). If this excess is attributed to charm production then 
one would get for the charm production cross section per nucleon $\sigma _c$:
\be
\sigma _c=(0.16\pm 0.023)\sigma _{pPb}/A_{Pb}.      \label{sig}
\ee
Here it is assumed that due to smallness of $\sigma _c$ the cross section
of charm hadroproduction on lead is proportional to the atomic number that 
is accounted by the rightmost factor $A_{Pb}=207$.
Another method of $\sigma _c$ estimate is based on the experimental fact that
some part (0.358) of cascades with the attenuation length less than 800 g/cm$^2$
have positive $b$. If this is ascribed to fluctuations in usual cascades without
charm particles then
\be
\sigma _c=0.5(1-0.358)\sigma _{pPb}/A_{Pb}=(0.321\pm 0.072)\sigma _{pPb}/A_{Pb}. \label{sigm}
\ee
Thus the charm production cross section per nucleon was estimated \cite{yak, viy} 
as $\sigma _c$=1.4 - 2.8 mb at energies 10 - 20 TeV. This is the very 
first estimate of the charm hadroproduction cross section from experimental
data at very high energies. Nowadays accelerator results are close to these
findings as discussed above (see Fig. 1).

\section{Theoretical interpretation}

\subsection{Analytical and computer solutions of the kinetic equations}

First, it was attempted to show by simple analytical means that production 
of the long-lived particles can give rise to the increase of the attenuation 
length with its subsequent decrease for those showers with increasingly high 
energies in which the charm particle escapes from installation before decay.
The simplified system of the kinetic equations consisted of three equation 
taking into account the evolution of the charm ($c$), nucleon ($N$) and pion 
($\pi ^{\pm }$) components:
\begin{eqnarray}
\frac {dS_c}{dz}=-\gamma S_c, \nonumber \\
\frac {dS_N}{dz}=-\beta S_N+\delta S_c, \nonumber  \\
\frac{dS_{\pi ^{\pm}}}{dz}=-\alpha S_{\pi ^{\pm}}+\frac {2}{3}\beta S_N+
\frac {2}{3}(\gamma -\delta )S_c      \label{syst} 
\end{eqnarray}
with the initial conditions
\begin{eqnarray}
S_c(0)=\frac {\sigma _c}{\sigma _t}\langle x_c\rangle, \nonumber  \\
S_N(0)=1-K_N, \nonumber                                             \\
S_{\pi ^{\pm}}(0)=\frac {2}{3}\left(K_N-
\frac {\sigma _c}{\sigma _t}\langle x_c\rangle \right ).
\label{inct} 
\end{eqnarray}
Here $S$ means the share of initial energy taken by the charm, nucleon 
and pion components, $z$ is the depth along the cascade 
axis, $\sigma _c/\sigma _t $ is the probability of the charm particle creation.
A single species of charm particles is considered here.

The coefficients in the equations are
\be
\alpha =\frac {1/3}{\lambda _{\pi }}, \;\; \beta =\frac {K_N}{\lambda _N},
\;\; \gamma =\frac {1}{\lambda _d}+\frac {K_c}{\lambda _c},
\;\; \delta =\frac {1-B}{\lambda _d},                    \label{abc}
\ee
where $\lambda _i, \;  K_i$ are the interaction lengths and inelasticity 
coefficients of the corresponding component, $\lambda _d$ is the decay 
length of the charm component, $B$ is the share of energy of pions in the 
decay of the charm particle to nucleon and pions.

This system has been solved. Here we however demonstrate only the qualitative 
features of the solution for the pion component energy with a particular choice 
$B=1, K_N=1$ because it can be written in the simple analytical form as
\be
\frac {3}{2}S_{\pi ^{\pm }}=e^{-\alpha z}+\frac {\sigma _c}{\sigma _t}
\frac {\langle x_c\rangle }{\gamma - \alpha }[\alpha e^{\-\alpha z}-
\gamma e^{-\gamma z}].   \label{anal}
\ee
The formula (\ref{anal}) clearly shows that the production of charm particles 
leads to a slight "deepening" in the cascade ionization curve (due to the term 
with the negative sign in the brackets) which, at higher energies, is 
replaced by the "hump" with a subsequent exponential decrease at larger depths.
This is the origin of the humps in the energy dependence of attenuation lengths.

After establishing the main qualitative feature in the behaviour of attenuation
lengths with the help of the simplified analytical approach the more rigorous
computer solutions of kinetic equations were attempted. The system of kinetic
equations looked like
\be
\left ( \frac {d}{dz}+\frac {1}{\lambda _i}+\frac {m_i}{E\tau _ic\rho }\right )
F_i(E,z)=\sum _j\frac {1}{\lambda _j}\int _E^{E_0} F_j(E',z)W_0(E,E')dE'+\sum _j
R_{ij}       \label{mcar}
\ee
for the energy distribution functions $F_i(E,z)$ of particles $i$ at 
the depth $z$. The outflow of particles $i$ due to interactions ($\lambda _i$)
and decay ($\tau _i$) is compensated by their production in inelastic 
interactions of particles $j$ ($W_{ij}$) and in decays ($R_{ij}$). 

Now four components were considered: both $\Lambda _c$ and $D$-mesons were
taken as charm particles. No accurate data about their lifetimes existed at 
the time when these calculations had been done. Therefore, the lifetime of 
$\Lambda _c$ varied from 1.7$\cdot 10^{-13}$ s to 3.5$\cdot 10^{-13}$ s
and the average $"D"$-lifetime was chosen as 6.3$\cdot 10^{-13}$ s.
The pion spectra were taken according to the standard CKP-prescription with
the experimentally known energy dependence of the mean multiplicity. The 
spectrum of produced $\Lambda _c$ was very hard, independent on $x$, i.e.,
$\langle x_{\Lambda _c}\rangle $=0.5 \cite{basi, bode, khod}. This is in 
charge for large $\gamma $-factors of $\Lambda _c$, i.e., for large decay
lengths. The $D$-spectrum was softer \cite{mcl, hwa} decreasing with $x$
as $1-x$ but somewhat harder in the case of primary pions to account for 
the leading effect. The interaction lengths were taken according to the 
additive quark model, i.e., they are almost equal for pions and $\Lambda _c$, 
1.5 times larger for nucleons and about twice smaller for $D$-mesons.
The inelasticity coefficients were put equal to $K_N$=0.63 \cite{baza}, 
$K_{\pi^{\pm }}$=0.7 \cite{baza}, $K_c$=0.1 \cite{khod, mcl}. The small
inelasticity coefficient for $\Lambda _c$ is crucial for their penetration
to larger depths compared to pions. Thus, the three factors of interaction
cross sections, decay lengths and inelasticity coefficients determine mostly
the increase of the attenuation lengths.
Electron-photon cascades  due to  decays of neutral pions 
were treated by common standards. The energy dependence of the charm production 
cross section was chosen so that it went through the experimental data at low 
ISR+FNAL energies (see Fig. 1) increasing to some constant values at very high 
energies. Its atomic number dependence was linear. 

The equations were solved 
with the initial conditions $F_i(0,E)=W_{iN}(E. E_0)$, i.e. with all cascades 
initiated at the same point in the calorimeter. The exponential fit of the 
behaviour of the cascade ionization with the distance in the calorimeter at 
the depths from 344 to 800 g/cm$^2$ shows the energy 
dependence of the attenuation lengths. It is depicted in Fig. 7 for variants
with high energy charm hadroproduction cross sections saturated at 2, 5, 10 mb 
at (and above) 100 TeV. The lines are marked by the corresponding numbers. 

If compared with experimentally measured values of attenuation lengths, the 
charm cross section about 5 mb would be preferred. However the peak is shifted 
to somewhat larger energies. It appears at proper position only in Monte Carlo
simulations taking into account the origin points of the individual cascades
and the correct estimation of energy released in the calorimeter.
The results for the standard cascade without charm production are shown by 
the dashed line. They are surely much below experimental data for all cascades
but agree well with the lower dots in Fig. 2 obtained for cascades with $b<0$.

Again, the qualitative effect of the increase and maximum in the energy
behaviour of attenuation lengths due to charm production has been demonstrated.
However, a single maximum appeared because, first, the two species of 
$D$-mesons were averaged with their lifetime chosen much closer to the 
$\Lambda _c$ lifetime compared to present values and, second,
no separation of cascades according to the sign of $b$ was attempted.
The very first interaction was always chosen at the top of the calorimeter.
No account of the energy spectrum and primary composition of cosmic rays was 
done. In principle, all these deficiencies can be cured by the proper Monte 
Carlo simulations.

\subsection{Monte Carlo simulations}

In the more detailed Monte Carlo cascade simulations \cite{drya, madi, dima}
some (but not all) of the above deficiencies were avoided. The positions of 
first interactions and values of energy release in the calorimeter were 
properly accounted. This has lead to better agreement with the energy 
locations of the maxima. Moreover, the structure appeared compared to Fig. 7.
Unfortunately, the charged and neutral $D$-mesons were again considered
together even though it was taken into account that the lifetimes of 
$\Lambda _c$ and $D$-mesons differ stronger than those values which were 
adopted above. This has also lead to some structure in 
the maximum of the attenuation lengths as shown by dots in Fig. 8. It reflects 
quite well the corresponding structure in experimental data (shown by 
crosses in Fig. 8) if no separation of 
cascades according to the sign of $b$ is done. The cascade simulation was 
performed for the energy region from 3 to 1000 TeV in intervals of
$\Delta \log E$=0.2 with an integral spectrum of primary cosmic ray hadrons
of a power law behaviour with an exponent equal to 1.8. For each step of 
$\Delta \log E$ a statistics of between 800 and 1600 cascades was required.
Charm production was only included for energies exceeding 300 GeV, below
which only standard cascades were allowed. The atomic number dependence
was slightly modified to take into account the screening in case of larger
charm cross section by replacing the linear dependence by 
$A^{1-\sigma _c/120 mb}$ which is really insignificant for small $\sigma _c$.
From these calculations and their comparison with experiment the value
of the charm hadroproduction cross section was estimated \cite{dyak, drya}
as 3 - 5 mb at energies about 100 TeV (see Fig. 1).

Unfortunately, Monte Carlo cascades were not separated according to the sign
of $b$. Namely this method has lead to three peaks in Fig. 2. Thus, we can 
not claim that the structure seen in Fig. 2 has been quantitatively described, 
even though Fig. 8 points in this direction. Nevertheless, the successful
description of the unseparated cascades is clearly seen. We conclude that at 
least qualitatively the specific structure in the energy behaviour of 
attenuation lengths has been understood.

\section{Conclusions and perspectives}

The long-flying cascades in the cores of extensive air showers have been
observed experimentally and explained theoretically as a result of charm
production at high energies. The peculiar structure in the energy 
dependence of the attenuation lengths has been ascribed to different species
of charm particles possessing different lifetimes and masses. 

There are several factors which determine the increase of the attenuation 
lengths. Charm particles are able to penetrate to large depths because they 
are rather heavy and spend less energy in inelastic collisions (small
inelasticity coefficients). Also, this is in charge of the leading effect
at their production and smaller interaction cross sections.

The main outcome of the analysis is the prediction of quite large cross sections 
for charm production which was done long before the accelerator data at high 
energies became available. These values looked suspiciously high if compared
to the data at lower energies. Now they are however confirmed by recent
experimental results of RHIC.

In view of better nowadays knowledge of the parameters of charm particles
and good progress in developing the models of charm hadroproduction, the next 
step in refining theoretical results is foreseen. Both improved Monte Carlo
simulations taking into account these developments and analysis of the simulated 
cascades by separating them into subgroups with the positive and negative signs
of their concentration near the cascade axis can be done. The quantitative
description of the three-maxima structure in Fig. 2 is the main goal. This is
especially important for understanding the energy spectra of charm particles
and more accurate estimates of their hadroproduction cross sections.

Charm particle decay can be important for EAS evolution in the atmosphere. It was 
pointed out in \cite{mcl} that the decay length of the charm particles becomes
comparable with the vertical size of the atmosphere at energies (1 - 7)$\cdot $
10$^{17}$ eV. Recently it was shown \cite{yatm} that the excess of EAS detected
by AGASA installation \cite{agas} at energies above the GZK cutoff can be 
related with the increasing role of charm particles. There exist some other
observations which could favour this interpretation. First, there is a strong
excess \cite{iek, niki} of showers with energies above $10^{18}$ eV at large 
zenith angles ($>60^o$). Second, experiments at Pierre Auger Observatory 
Surface Array revealed that "a significant number of very horizontal events 
are detected, offering a novel view of EAS" \cite{bert}.

The experiment for confirming the crucial role of charm particles for hadronic
cascades was proposed in \cite{dmy}. The calorimeter with an air gap about
2.5 m should be used. Then the charm particles with energies less than 70 TeV,
produced in the upper rows of the calorimeter, decay inside this air gap.
That would intensify the energy release in the upper rows of the lower part 
of the calorimeter, i.e., lead to a hump in the energy distribution. The 
analogous experiment was proposed \cite{sstr} with X-ray films. Such 
experiment is under way in Tien Shan station of LPI now. \\

{\bf Acknowledgements}

This work has been supported in part by the RFBR grants 03-02-16134, 
04-02-16320, 04-02-16445, NSH-1936.2003.2.\\

{\bf Figure captions.}

\begin{tabbing}
Fig. 1. \= The compilation of the charm hadroproduction cross sections at \\
        \> different energies from accelerator and cosmic ray data.\\
Fig. 2. \> The attenuation length distributions for cascades with opposite \\
        \> signs of the derivative of the concentration $b$. The low-lying \\
        \> points correspond to the "standard" cascades with $b<0$. The upper\\
        \> points are due to the long-flying cascades with $b>0$, enriched \\
        \> by charm particles. The three-maxima structure is clearly seen. \\
Fig. 3. \> The cascade at the energy 37.6 TeV averaged over 765 events. \\
Fig. 4. \> The correlation between the derivative of the concentration $b$ \\
        \> and the attenuation lengths of the cascades.\\
Fig. 5. \> The distribution of the main secondary maxima of the ionization \\
        \> curves for the two groups of cascades. Squares are for cascades \\
        \> with $L<800$ g/cm$^2$, crosses - for $L>800$ g/cm$^2$.\\
Fig. 6. \> The cascades distributions for different attenuation lengths \\
        \> in groups with $b<0$ (solid line) and $b>0$ (dashed line).\\
Fig. 7. \> The energy dependence of the attenuation lengths obtained from the\\
        \> computer solution of the kinetic equations for different "asymptotic"\\
        \> values (shown by numbers in mb) of the charm production cross sections.\\
Fig. 8. \> The energy dependence of the attenuation lengths from the experiment\\
        \> (circles) and Monte Carlo computer simulations (squares) for all cascades\\
        \> without separating them into two subsamples according to the sign of $b$.\\
\end{tabbing}

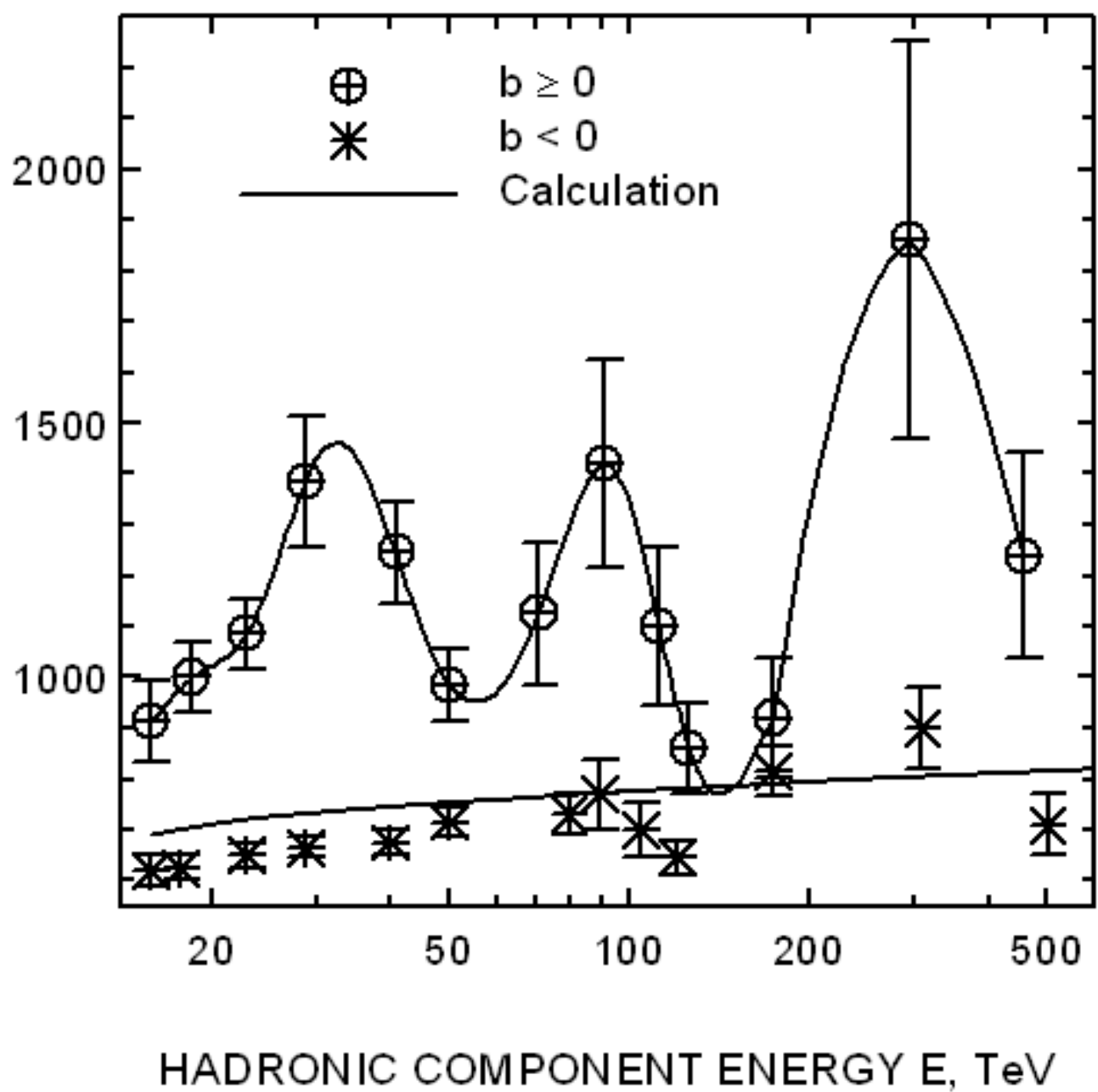
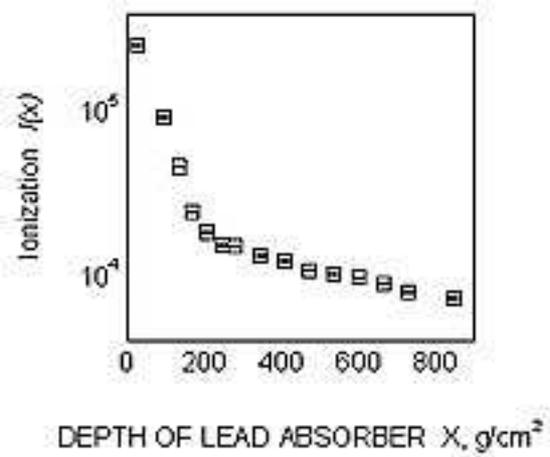
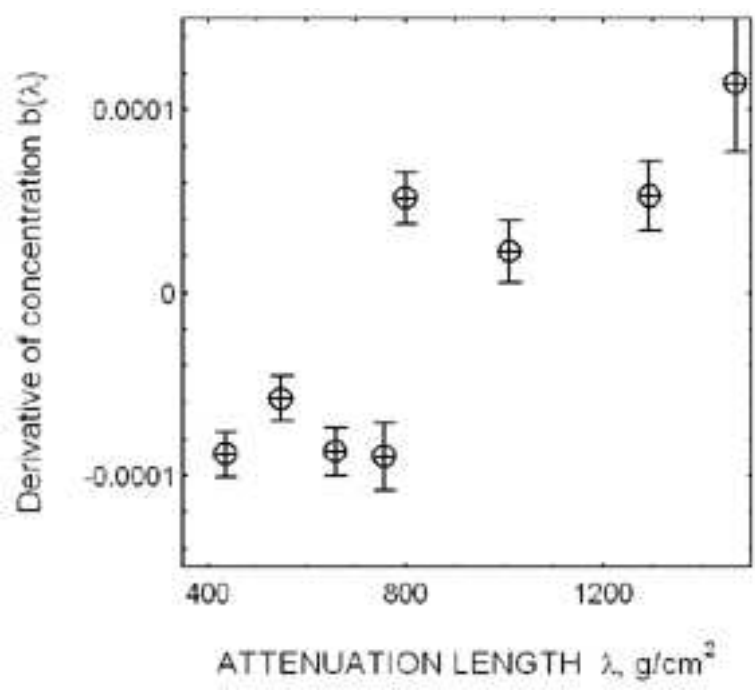
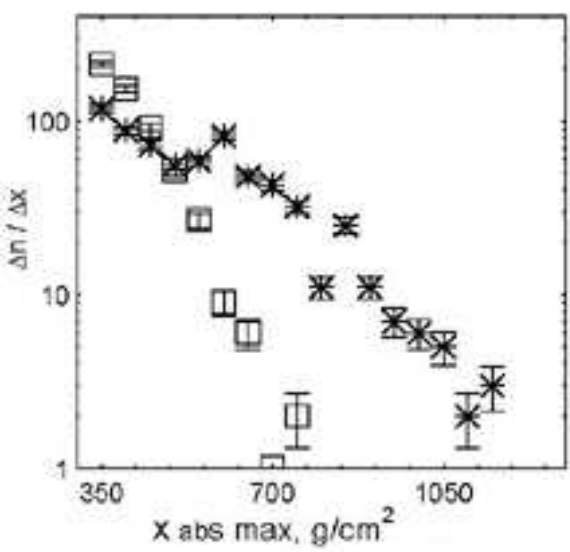
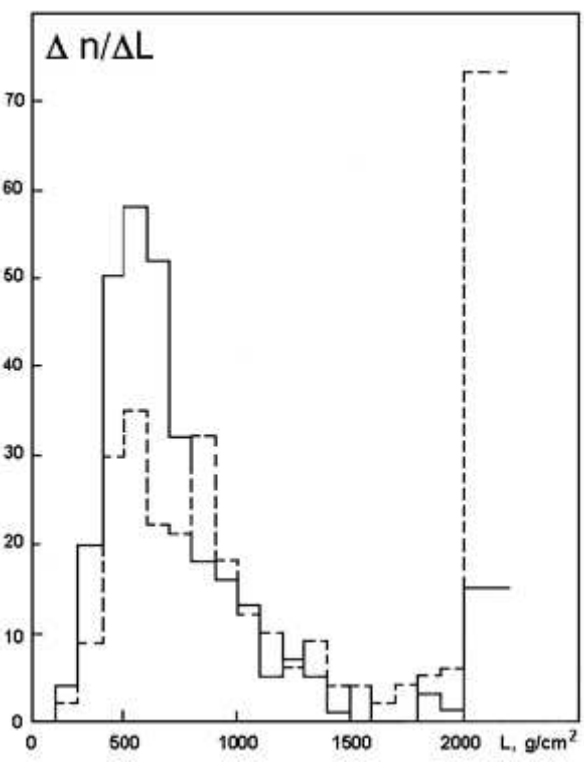
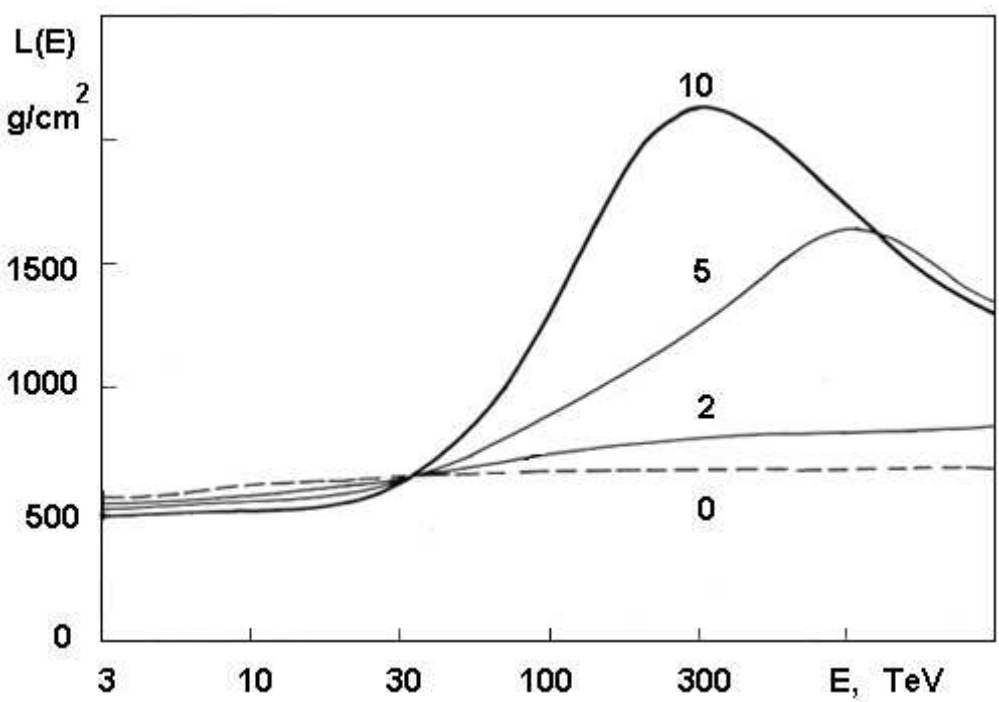
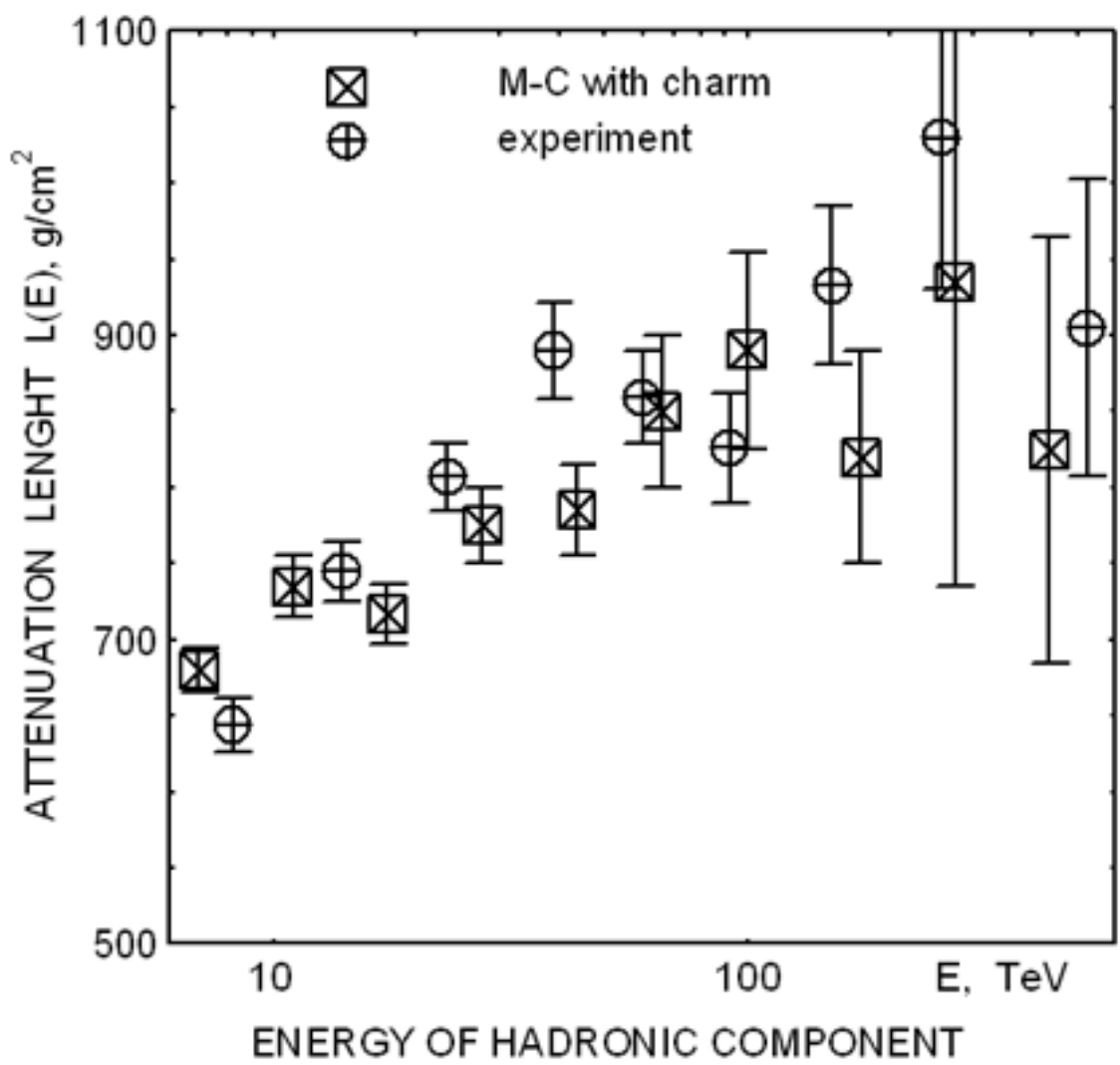
\end{document}